\documentstyle[12pt]{article}
\newcommand{\vf}{\varphi}
\newcommand{\be}{\begin{equation}}
\newcommand{\ee}{\end{equation}}
\newcommand{\ba}{\begin{eqnarray}}
\newcommand{\ea}{\end{eqnarray}}
\newcommand{\no}{\nonumber\\}
\newcommand{\bi}{\bibitem}
\begin{document}

\thispagestyle{empty}
\begin{flushright}
IFA-FT-397-1994 \\
July 1994
\end{flushright}
\bigskip\bigskip\begin{center}
{\bf \Huge{GENERALIZED}}
\end{center}
\begin{center}
{\bf \Huge{RUNGE - LENZ VECTOR}}
\end{center}
\begin{center}
{\bf \Huge{IN TAUB -NUT SPINNING SPACE}}
\end{center}
\vskip 1.0truecm
\centerline{{\bf\Large
{Mihai Visinescu\footnote{E-mail address:~~ MVISIN@ROIFA.BITNET
{}~~~~or~~~~ MVISIN@IFA.RO}}}}
\vskip5mm
\centerline{Department of Theoretical Physics}
\centerline{Institute of Atomic Physics, P.O.Box MG-6, Magurele,}
\centerline{Bucharest, Romania}
\vskip 2cm
\bigskip \nopagebreak \begin{abstract}
\noindent
The generalized Killing equations and the symmetries of Taub-NUT spinning
space are investigated. For spinless particles the
Runge-Lenz vector defines a constant of motion directly, whereas for
spinning particles it now requires a non-trivial contribution from spin.
The generalized Runge-Lenz vector for spinning Taub-NUT space is completely
evaluated.
\end{abstract}

{}~~~~~~11.30.-j - Symmetry amd conservation laws

\newpage\setcounter{page}1
\section{Introduction}
Spinning particles, such as Dirac fermions, can be described by
pseudo- classical mechanics models involving anti- commuting c- numbers for
the spin-degrees of freedom. The configuration space of spinning particles
(spinning space)is an extension of an ordinary Riemannian manifold,
parametrized by local co-ordinates {$\{$}$x^\mu${$\}$}, to a graded manifold
parametrized by local co-ordinates {$\{$}$x^\mu, \psi^\mu${$\}$}, with the
first set of variables being Grassmann-even ( commuting ) and the second set
Grassmann-odd (anti-commuting) [1-9].

In this paper we investigate the Runge-Lenz vector for the motion of
pseudo-classical spinning point particles in the Euclidean Taub-NUT space
[10]. The Kaluza-Klein monopole was obtained by embedding the Taub-NUT
gravitational instanton into five-dimensional theory, adding the time
co-ordinate in a trivial way [11, 12]. Its line element is expressed as:
\ba
ds^2_5&=&-dt^2+ds^2_4\no
&=&-dt^2+V^{-1}(r)[dr^2+r^2d\theta^2+r^2\sin^2\theta\,d\vf^2]\no
& &+V(r)[dx^5+\vec{A}(\vec{r})\,d\vec{r}\,]^2
\ea
where $\vec{r}$ denotes a three-vector $\vec{r} =(r, \theta, \vf)$ and
the gauge field $\vec{A}$ is that of a monopole
\ba
A_r=A_\theta&=&0,~~~A_{\vf}=4m(1-\cos\theta)\no
\vec{B}&=&rot\vec{A}={4m\vec{r}\over r^3} .
\ea

The function $V(r)$ is
\be
V(r)=\left(1+{4m\over r}\right)^{-1}
\ee
and the so called NUT singularity is absent if $x^5$ is periodic with
period $16\pi m$ [13].

It is convenient to make the co-ordinate transformation
\be
4m(\chi+\vf)=-x^5
\ee
with $0\leq \chi < 4\pi$, which converts the four-dimensional line
element $ds_4$ into
\ba
ds^2_4&=&V^{-1}(r)[dr^2+r^2d\theta^2+r^2\sin^2\theta\, d\vf^2]
+16m^2 V(r)[d\chi+\cos\theta\, d\vf]^2 \no
&=& g_{\mu\nu}~dx^{\mu}dx^{\nu}.
\ea
Remarkably, the same object has re-emerged in the study of monopole
scattering. Slow Bogomolny-Prasad-Sommerfield monopoles move along
geodesics in a four-dimensional curved space with the line element $ds_{4}$
 [14-16].

The geodesic motion of a spinless particle of unit mass in (5) can be
derived from the action:
\be
 {\cal S}=\int_{a}^{b}d\tau \,{1\over 2}\,g_{\mu \nu}(x)\,\dot{x}^\mu
\,\dot{x}^\nu\,  .
\ee
Here and in the following the overdot denotes an ordinary proper-time
derivative $d/d\tau$.

The invariance of the metric (5) under spatial rotations and $\chi$
translations is generated by four Killing vectors
\be
D^{(\alpha)}=R^{(\alpha)\mu}\,\partial_\mu,~~~~\alpha=1,\cdots ,4
\ee
where
\ba
D^{(1)}&=&-\sin\vf\,{\partial\over \partial\theta}-\cos\vf\,\cot\theta
\,{\partial\over\partial\vf}+{\cos\vf\over\sin\theta}\,{\partial\over
\partial\chi}\no
D^{(2)}&=&\cos\vf\,{\partial\over \partial\theta}-\sin\vf\,\cot\theta
\,{\partial\over\partial\vf}+{\sin\vf\over\sin\theta}\,{\partial\over
\partial\chi}\no
D^{(3)}&=&{\partial\over\partial\vf}\no
D^{(4)}&=&{\partial\over\partial\chi}.
\ea
In the purely bosonic case these invariances would correspond to
conservation of angular momentum and "relative electric charge" [16-18]:
\be
\vec{j}=\vec{r}\times\vec{p}\,+\,q\,{\vec{r}\over r} .
\ee
\be
q=16m^2\,V(r)\,(\dot\chi+\cos\theta\,\dot\vf)
\ee
where
\be
\vec{p}={1\over V(r)}\dot{\vec{r}}
\ee
is the "mechanical momentum" which is only part of the momentum
canonically conjugate to $\vec{r}$. Energy, given by
\be
E={1\over 2}\, g^{\mu\nu}\,\Pi_\mu\,\Pi_\nu\,
={1\over 2}V^{-1}(r)\left[\dot{\vec{r}}^{\,2} +\left( {q\over
4m}\right)^2\right]
\ee
is also conserved , $\Pi_\mu$ being the covariant momentum.

Finally, there is a conserved vector analogous to the Runge- Lenz vector
of the Coulomb problem:
\ba
\vec{{\cal K}}&=&\vec{K}_{\mu\nu}\dot{x}^\mu\dot{x}^\nu\no
&=&\vec{p}\times\vec{j}\, +\, \left({q^2\over 4m}-4mE\right)
{\vec{r}\over r}.
\ea
Its existence is rather surprising in view of the complexity of motion
in the Taub- NUT space [16-19].

These results, combined, imply that the trajectories are conic sections
[16-18].
\section{Motion in spinning space}
The extension of the Euclidean Taub- NUT space with additional
fermionic dimensions, parametrized by vectorial Grassmann co-ordinate
{$\{$}$\psi^\mu${$\}$} follows naturally. An action for the geodesics
of spinning space is
\be
{\cal S}=\int_{a}^{b}d\tau \left(\,{1\over 2}\,g_{\mu \nu}(x)\,\dot{x}^\mu
\,\dot{x}^\nu\, +\, {i\over 2}\, g_{\mu \nu}(x)\,\psi^\mu \,{D\psi^\nu\over
D\tau} \right)
\ee
where the covariant derivative of $\psi^\mu$ is defined by
\be
{D\psi^\mu\over D\tau}=\dot{\psi}^\mu+\dot{x}^\lambda \,\Gamma^\mu
_{\lambda\nu}\,\psi^\nu .
\ee

The concept of Killing vector can be generalized to the case of spinning
manifolds. For this purpose it is necessary to consider variations of
$x^\mu$ and $\psi^\mu$ which leave the action ${\cal S}$ invariant modulo
boundary terms :
\be
\delta {\cal S}=\int_{a}^{b} d\tau {d\over d\tau}\left(\delta x^\mu\, p_\mu
-{i\over 2}\, \delta \psi^\mu \,g_{\mu\nu}\, \psi^\nu- {\cal J}(x, \dot x,
\psi)\right)
\ee
where $p_\mu$ is the canonical momentum conjugate to $x^\mu$
\be
p_\mu=g_{\mu\nu}\dot{x}^\nu+{i\over 2}\Gamma_{\mu\nu;\,\lambda}
\,\psi^\lambda\psi^\nu=\Pi_\mu + {i\over 2}\Gamma_{\mu\nu;\,\lambda}
\,\psi^\lambda\psi^\nu .
\ee

{}From Noether's theorem, if the equations of motion are satisfied,
the quantity ${\cal J}$ is a constant of motion.

If we expand ${\cal J}$ in a power series in the covariant momentum
\be
{\cal J}(x, \dot x, \psi) ={\cal J}^{(0)}(x, \psi) +
\sum_{n=1}^{\infty}{1\over n!}\Pi^{\mu_1}\cdots \Pi^{\mu_n}
{\cal J}^{(n)}_{\mu_1 \cdots \mu_n}(x, \psi)
\ee
then ${\cal J}$ is a constant of motion if its components satisfy a
generalization of the Killing equation [6, 7] :
\be
{\cal J}^{(n)}_{(\mu_1 \cdots \mu_{n};\,\mu_{n+1})}+ {\partial
{\cal J}^{(n)}_{(\mu_1\cdots \mu_n}\over\partial\psi^\sigma}
\Gamma^{~~~~~~\sigma}_{\mu_{n+1})\lambda}\ \psi^\lambda=
{i\over 2}\,\psi^\sigma \,\psi^\lambda\, R_{\sigma\lambda\nu(\mu_{n+1}}
{\cal J}^{(n+1)\nu}_{\mu_1\cdots\mu_n)} .
\ee

In general the symmetries of a spinning-particle model can be divided
into two classes.  First, there are four independent {\it generic}
symmetries  which exist in any theory [6,7] :

\begin{enumerate}
\item{Proper-time translations and the corresponding constant of
motion is the energy $E$ (12)}
\item{Supersymmetry generated by the supercharge
\be
Q=\Pi_\mu\,\psi^\mu
\ee}
\item{Chiral symmetry generated by the chiral charge
\be
\Gamma_{*}={1\over 4!}\sqrt{g}\epsilon_{\mu\nu\lambda\sigma}
\,\psi^\mu\,\psi^\nu\,\psi^\lambda\,\psi^\sigma
\ee}
\item{Dual supersymmetry, generated by the dual supercharge
\be
Q^{*}={1\over 3!}\,\sqrt{g}\epsilon_{\mu\nu\lambda\sigma}
\,\Pi^\mu\,\psi^\nu\,\psi^\lambda\,\psi^\sigma  .
\ee}
\end{enumerate}

The second kind of conserved quantities, called {\it non-generic},
depend on the explicit form of the metric $g_{\mu\nu}(x)$. Let us
examine in detail the generalized Killing Eq.(19) for the specific case
of Taub- NUT space. For this purpose we shall write the power expansion
(18) in the form
\be
{\cal J}(x, \dot x, \psi) =B + R_{\mu}\Pi^{\mu} + {1\over 2!}K_{\mu\nu}
\Pi^{\mu}\Pi^{\nu} + {1\over 3!}L_{\mu\nu\lambda}\Pi^{\mu}\Pi^{\nu}
\Pi^{\lambda}+\cdots
\ee
with some notations similar to the ones used in eqs.(7) and (13). The
generalized Killing eq.(19) reduces for the lowest compoments to
\ba
&~&B_{,\mu} + {\partial B\over \partial \psi^\sigma}\Gamma^\sigma_{\mu
\kappa}\psi^\kappa = {i\over 2}\psi^\rho\psi^\sigma R_{\rho\sigma\kappa\mu}
R^\kappa\\
&~&R_{(\mu ;\nu)} + {\partial R_{(\mu} \over \partial \psi^\sigma} \Gamma
^\sigma _{\nu)\kappa}\psi^\kappa = {i\over 2} \psi^\rho\psi^\sigma
R_{\rho\sigma\kappa(\mu}K^{~\kappa}_{\nu)}\\
&~&K_{(\mu\nu ;\lambda)} + {\partial K_{(\mu\nu}\over \partial\psi^\sigma}
\Gamma^\sigma_{\lambda)\kappa}\psi^\kappa={i\over 2 }\psi^\rho\psi^\sigma
R_{\rho\sigma\kappa(\mu}L^\kappa_{\nu\lambda)}.
\ea

The purely bosonic ($\psi$ - independent) parts of these equations
reduces to the Killing equations for the isometries of the usual Taub-
NUT space discussed in the previous chapter. The first equation (24)
implies that $B$ is an irrelevant  constant, while the second equation
(25) is the standard equation for the Killing vectors $R^{(\alpha)}_{\mu}
{}~(\alpha\,=\,1,...,4)$. The third equation (26) will be analysed in the
next section.

The first generalized Killing equation (24) shows that with each
Killing vector $R^{(\alpha)}_\mu$ there is an associated Killing
scalar $B^{(\alpha)}$. This equation  has been solved in Ref.[20] for
the spinning Taub- NUT space using for the Killing vectors
$R^{(\alpha) }_\mu$ the expressions (7) and (8). We get the constants
of motion for the spinning Taub- NUT space in the form:
\ba
{\vec J}&=& {\vec B} + {\vec j}\\
J^{(4)} &=& B^{(4)} + q
\ea
where ${\vec J} =(J^{(1)}, J^{(2)}, J^{(3)})$ and ${\vec B} =
(B^{(1)}, B^{(2)}, B^{(3)})$. Eq.(27) shows that the conserved total
angular momentum is the sum of the angular momentum ${\vec j}$ (9) and
the spin angular momentum contribution described by ${\vec B}$.
Similarly, from eq.(28), we conclude that the "relative electric
charge" $q$ (10) is no longer conserved and the new conserved quantity
$J^{(4)}$ received a spin dependent part.
\section{Runge- Lenz vector in spinning Taub- NUT space}
The Runge- Lenz vector ${\vec K} = ( K^{(1)}, K^{(2)}, K^{(3)})$ given
in eq.(13) is quadratic in 4- velocities in the usual Taub- NUT space
and the 3- Killing tensors [21] $K^{(\alpha)}_{\mu\nu}~~(\alpha=1, 2,
3)$ satisfy a generalized Killing equation
\be
K^{(\alpha)}_{(\mu\nu;\lambda)}  = 0~~~~\alpha=1, 2, 3
\ee
which is exactly the purely bosonic  part of eq.(26). Eq.(29) can be
verified explicitely and it was analysed in detail in [17] where it is
shown the origin of these extra-conserved quantities for the geodesic
motions in the Taub- NUT space. The existence of these 3- Killing
tensors (Stackel- Killing tensors) can be related to the existence on
Taub- NUT space of a Killing- Yano 2- form $f_{\mu\nu}=-f_{\nu\mu} $
such that [22]
\be
K^\mu_{~\nu}=f^\mu_{~\lambda}f^\lambda_{~\nu}
\ee

A similar situation is found in the Kerr- Newman solutions of the
combined Einstein- Maxwell equations [23-28]. The extension of these
results to the spinning Kerr- Newman space is done in [9].

In this section we shall evaluate the generalized Runge- Lenz vector
for the Taub- NUT space. For this purpose we shall solve eq.(25) using
in the r.h.s the Killing tensor components which appear in (13). The
complete Killing vectors can be written in the following form:
\be
{\cal R}^{(\alpha)}_{\mu}=R^{(\alpha)}_{\mu}+S^{(\alpha)}_{\mu}
{}~~~~,\alpha =1,2,3
\ee
where $R^{(\alpha)}_{\mu}$ are known from the scalar case,eqs.(7),(8)
and they correspond to the angular momentom (9). The $\psi$ - dependent
parts of the Killing vectors $S^{(\alpha)}_{\mu}$ contribute to
the Runge- Lenz vector for the spinning space
\be
\vec{{\cal K}} =\vec{K}_{\mu\nu}\cdot\dot{x}^\mu\dot{x}^\nu + \vec{S}
_\mu \cdot\dot{x}^\mu
\ee

The evaluation of the last term involves some long and tedious
calculations. The results are the following:
\newpage
\ba
\lefteqn{S^{(1)}_\mu \cdot \dot{x}^\mu=}\no
& &\left[-{1\over 2}(4m+r)\cos\theta\cos\vf\cdot
S^{r\theta}+\right. \no
& &\left. 2mr\cos\theta\sin\vf\cdot S^{\theta\vf}
+2mr\sin\vf\cdot S^{\theta\chi}+\right.\no
& &\left. {1\over 2}(4m+r)\sin\theta\sin\vf\cdot S^{r\vf}+
2mr\sin\theta\cos\theta\cos\vf \cdot S^{\vf\chi}\right] \dot{r}+\no
& &\left[{1\over 2}r(4m+r)\sin\theta\cos\vf\cdot S^{r\theta}
-{2mr(6m+r)\over 4m+r}\cos\theta\sin\vf\cdot S^{r\vf}-\right.\no
& &\left.{2mr(6m+r)\over 4m+r}sin\vf\cdot S^{r\chi}+ \right.\no
& &\left.{1\over 2}r^2(6m+r)\sin\theta\sin\vf S^{\theta\vf}-
2mr^2\sin^2\theta\cos\vf S^{\vf\chi}\right] \dot{\theta}+\no
& &\left[{2mr(6m+r)\over 4m+r}\cos\theta\sin\vf S^{r\theta}+\right.\no
& &\left({1\over 2}r(4m+r)\sin^3\theta\cos\vf+
{128m^4r\over(4m+r)^3}\sin\theta\cos^2\theta\cos\vf\right) S^{r\vf}-\no
& &\left(2mr+{4m^2r\over 4m+r}-{128m^4r\over(4m+r)^3}\right)\sin\theta
\cos\theta\cos\vf S^{r\chi}+\no
& &\left({r^2(32m^3+64m^2r+14mr^2+r^3)\over 2(4m+r)^2}\sin^2\theta\cos
\theta\cos\vf+\right.\no
& &\left. {16m^3r^2\over (4m+r)^2}\cos^3\theta\cos\vf\right) S^{\theta
\vf}+\no
& & \left(2mr^2\sin^2\theta\cos\vf+{16m^3r^2\over (4m+r)^2}\cos^2\theta
\cos\vf\right) S^{\theta\chi}-\no
& &\left.{16m^3r^2\over (4m+r)^2}\sin\theta\cos\theta\sin\vf S^{\vf\chi}
\right]\dot{\vf}+\no
& &\left[ {2mr(6m+r)\over 4m+r}\sin\vf S^{r\theta}+{128m^4r\over
(4m+r)^3}\sin\theta\cos\vf S^{r\chi}\right.+\no
& &\left(2mr +{4m^2r\over 4m+r} +{128m^4r\over (4m+r)^3}\right)\sin\theta
\cos\theta\cos\vf S^{r\vf}+\no
& &\left({16m^3r^2\over (4m+r)^2}\cos^2\theta\cos\vf -\right.\no
& &\left.{64m^3r^2+16m^2r^3+2mr^4\over (4m+r)^2}\sin^2\theta\cos\vf\right)
S^{\theta\vf}+\no
& &\left.{16m^3r^2\over (4m+r)^2}\cos\theta\cos\vf S^{\theta\chi}-
{16m^3r^2\over(4m+r)^2}\sin\theta\sin\vf S^{\vf\chi}\right]\dot{\chi}
\ea
\newpage
\ba
\lefteqn{S^{(3)}_\mu \cdot \dot{x}^\mu=}\no
& &\left[{1\over 2}(4m+r)\sin\theta S^{r\theta}-2mr\sin^2\theta S^{\vf\chi}
\right]\dot{r}+\no
& &\left[{1\over 2}r(4m+r)\cos\theta S^{r\theta}-2mr^2\sin\theta\cos\theta
S^{\vf\chi}\right]\dot{\theta}+\no
& &\left[\left({128m^4r\over (4m+r)^3}\cos^3\theta+{1\over 2}r(4m+r)
\sin^2\theta\cos\theta\right)S^{r\vf}\right.+\no
& &\left({128m^4r\over (4m+r)^3}\cos^2\theta+{2mr(6m+r)\over 4m+r}
\sin^2\theta\right) S^{r\chi}-\no
& &\left({1\over 2}r^2(6m+r)\sin^3\theta +{48m^3r^2\over (4m+r)^2}
\sin\theta\cos^2\theta\right) S^{\theta\vf}+\no
& &\left. mr^2\left(2-{16m^2\over (4m+r)^2}\right)\sin\theta\cos\theta
S^{\theta\chi}\right]\dot{\vf}+\no
& &\left[\left(-2mr\sin^2\theta-{4m^2r\over 4m+r}\sin^2\theta
+{128m^4r\over (4m+r)^3}\cos^2\theta\right) S^{r\vf}+\right.\no
& &{128m^4r\over (4m+r)^3}\cos\theta S^{r\chi}-
{16m^3r^2\over (4m+r)^2}\sin\theta S^{\theta\chi}-\no
& &\left.2mr^2\left(1 + {24m^2\over (4m+r)^2}\right)\sin\theta\cos\theta
S^{\theta\vf}\right]\dot{\chi}
\ea

The component $S^{(2)}_\mu\cdot\dot{x}^\mu$ can be obtain from
$S^{(1)}_\mu\cdot\dot{x}^\mu$, eq.(33), using the substitutions :
\ba
\sin\vf\longrightarrow &-&\cos\vf\no
\cos\vf\longrightarrow & &\sin\vf
\ea

In eqs.(33),(34) we used the quantity
\be
S^{\mu\nu}=-i\psi^\mu \psi^\nu
\ee
which can be regarded as the spin- polarization tensor of the particle [1-9].
\section{Concluding remarks}
The main purpose of this work has been the evaluation of the Runge-
Lenz vector for the spinning Taub- NUT space. Combining this result with
those from Ref.[20] we have a complete description of the motion of
spinning particles in Taub- NUT space.

Unfortunately the formulae are quite intricate as compared to the
scalar case ( such as eqs. (9), (13)). We mention that similar involved
formulae appear also in the study of the Schwarzchild [8] and Kerr-
Newman [9] spinning spaces.

Extension of the analysis of the hidden symmetries of Taub- NUT space
from Ref. [17] for the spinning case is possible and necessary. In
general it is desirable to have a deeper understanding of the role of
the Runge- Lenz vector for the motion of spinning particles. Having
described the dynamical symmetries for the classical problem, the next
step is to go into the quantum mechanical picture.

These problems are under investigations.
\subsubsection*{Acknowledgements}
The author have benefitted from correspondence with van Holten
regarding the Killing- Yano tensors. This work has been completed
during a visit to the Department of Physics, Technion, Haifa. The
author is grateful to M.Moshe for making this visit possible and for
useful discussions.

\end{document}